# Optimization of Dielectric Rod Antenna Design in Millimeter Wave Band for Wireless Communications

Nancy Ghattas[1], Atef M. Ghuniem[2], Sherif M. Abuelenin[3]

[1]Faculty of Engineering, Suez Canal University
nancyghattas@hotmail.com
[2]Faculty of Engineering, Suez Canal University
atghuniem@yahoo.com
[3]Faculty of Engineering, Port-Said University
s.abuelenin@eng.psu.edu.eg

**ABSTRACT**

Dielectric rod antennas are promising candidates to be used in millimeter wave band wireless communications. The optimization of a patch-fed dielectric rod antenna design at millimeter wave band is presented in this work. This design is composed of simple dielectric rod antenna made of low cost Teflon with relative permittivity of 2.1 that is fed by a rectangular patch antenna. Compared to the conventionally used tapered dielectric rod antennas, this configuration is capable of producing higher gain for a given antenna length. The design is optimized to achieve gain up to 17.5 dBi and impedance bandwidth up to 4.4 GHz at 60 GHz. A 4-element array is proposed to increase the gain up to 21 dBi. Simulations are performed using CST microwave studio, and their results are presented.

## I. INTRODUCTION

In recent years, the need for high data rate communication increased for indoor as well as outdoor applications and multimedia services, e.g., HDMI and HDTV, which requires high channel capacity. Studies show that an efficient solution lies in offering large bandwidth, which is not available in the low frequency spectrum, e.g., at 2 GHz. Therefore, the mm-wave spectrum at 60 GHz is of high interest not only because it is a license-exempt frequency band, but it also provides a vast bandwidth of 4 GHz and more. Also, there is an increasing demand of high gain antennas used for consumer devices. Some approaches to increase the antenna gain at 60 GHz have been reported in [1] and [2], though high gain performance can be achieved, but the relatively large dimension and the increased complexity of the complete antenna system may limit its use. The desired antenna has to be compatible with integrated circuits, and should have the benefits of small size and low cost.

Both dielectric rod antennas and dielectric resonator antennas have been researched and studied for many years. Resonator antenna has lower gain comparing with tapered rod antenna because most of the wave inside the antenna is standing wave. There are two shapes of dielectric rod antenna. One is the circular cross section antenna, and the other is the rectangle cross section antenna. The circular cross section dielectric rod is considered in this paper. Dielectric rod antennas have also the advantages of small size and light weight. Numerical analysis of dielectric rod antennas has been developed using finite difference time-domain (FDTD) method [3]. The dielectric rod antenna belongs to the family of surface wave antennas. The radiation mechanism of dielectric rod antenna can be explained by the discontinuity-radiation concept [4]. The increase in popularity of the dielectric rod with circular cross section is because of its wide bandwidth, shape, ability to create a symmetric radiation pattern, low polarization cross coupling, ease of fabrication, and low cost [5]-[7].

This paper presents a design optimization of a tapered dielectric rod, shown in Fig. 1. This dielectric rod is mainly excited by a patch antenna and a circular waveguide as shown in Fig. 2.

To improve the radiation pattern using dielectric rod antennas, a combination of tapered and cylindrical dielectric sections is used.

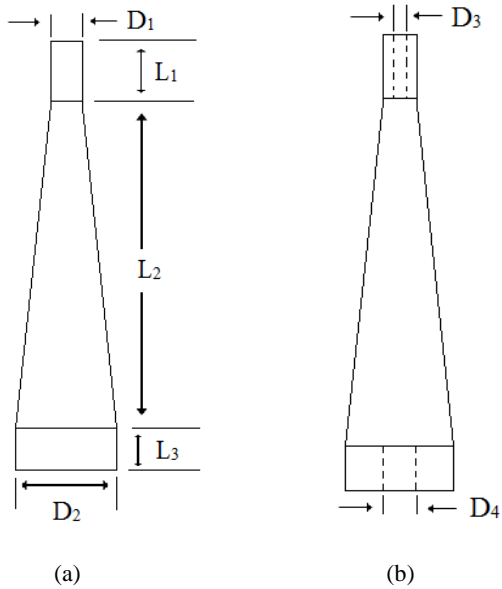

Fig. 1. Dielectric rod antenna geometry (a) Solid uniform parts, (b) Hollow uniform parts for single rod.

## II. THEORY OF DIELECTRIC ROD ANTENNA

The operation concept of the antenna is based on the effect of transformation of the dielectric waveguide surface waves in the volume waves. An axial surface wave is a plane wave that propagates in the axial direction of a cylindrical interface of two different media without radiation. The axial surface waves are plane waves because the phase remains constant along a plane perpendicular to the cylinder axis. They are also inhomogeneous because the field is not constant along surfaces of constant phase. The hybrid $HE_{11}$ mode is the dominant surface wave mode and is used most often with dielectric rod antennas. The higher, transversal modes $TE_{01}$ and $TM_{01}$ produce a null in the end-fire direction or are below cut-off. The $HE_{11}$ mode is a slow wave when the losses in the rod material are small. In this case, increasing the rod diameter will result in an even slower $HE_{11}$ surface wave of which the field is more confined to the rod [4]. The taper can be then introduced here to reduce the reflected wave.

This taper improves the matching of the rod antenna to free-space. Therefore, this taper is often called an impedance matcher between the wave impedance inside the rod to the wave impedance in free-space. Within the usable frequency range, the refracted wave combines with the guided surface wave to focus the radiated field and achieve a higher gain. Therefore, a choice of an optimal rod length is important since a very long rod may cause destructive additions of refracted and guided waves [8].

## III. ANTENNA DESIGN

The dielectric rod antenna is designed using CST simulator. The overall configuration consists of three parts: patch antenna as the feeding part, dielectric rod and circular waveguide. The proposed structure of dielectric rod is shown in Fig. 1. The tapered section is to reduce the Side Lobe Level (SLL), and the uniform section is to maximize gain. The maximum end-fire radiation is obtained by adjusting the length of those sections.

Since the surface wave radiates only at discontinuities, the total pattern of this antenna is formed by interference between the feed and terminal patterns. The feed couples a portion of the input power into a surface wave, which travels along the antenna structure to the termination, where it radiates into space. The ratio of power in the surface wave to the total input power is called the efficiency of excitation. The power not coupled into the surface wave is directly radiated by the feed in a pattern resembling that radiated by the feed when no antenna structure is in front of it.

The radiation characteristics of a dielectric rod antenna mainly depend on the dielectric material used to fabricate the antenna, its physical structure, and the excitation method used to feed the antenna. Hence the design criteria of dielectric rod antennas include the selection of dielectric material with proper dielectric constant, provision for exciting the dielectric rod in $HE_{11}$ mode, and the determination of diameter, length, and taper angle of the rod.

The dielectric rod antennas are tapered until the impedance of the antenna becomes equal to that of free space. Tapering the rod minimizes reflections at the free end of the rod.

## A. Designed dielectric rod antenna:

Many design iterations have been done on dielectric rod with different shapes. Optimized gain up to 17.5 dBi resulted with tapered circular dielectric rod. This rod is made of low cost Teflon with relative permittivity of 2.1. The detailed dimensions of the rod are shown in Fig. 1a, where $L_1=\lambda_0$, $L_2=6\lambda_0$, $L_3=0.6\lambda_0$, $D_1=0.28\lambda_0$ and $D_2=0.6\lambda_0$. The dielectric rod antenna consists of top and bottom solid uniform sections and a middle tapered section. For increasing impedance bandwidth, the solid uniform sections are replaced with hollow parts as shown in Fig. 1b where diameter for hollow parts are $D_3=0.08\lambda_0$ and $D_4=0.28\lambda_0$. This suggested hollow structure has the advantage of increasing the bandwidth from 2.2 to 4.4 GHz approximately.

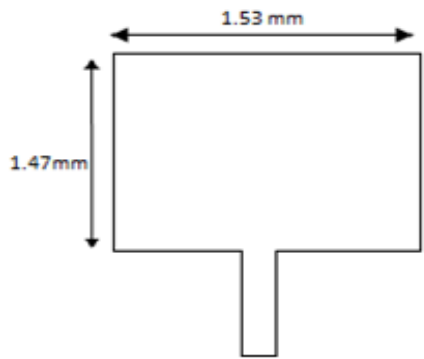

(a)

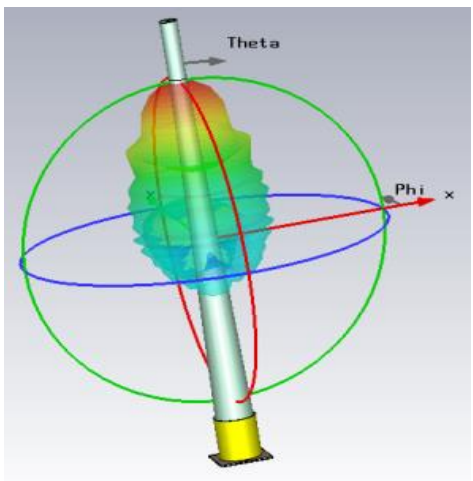

(b)

Fig. 2. (a) Single patch antenna, (b) Dielectric rod antenna fed by patch and surrounded by circular waveguide.

## B. Feeding part:

The dielectric rod is fed by a simple rectangular patch antenna with dimensions of 1.53 mm (length) and 1.49 mm (width) approximately as shown in Fig. 2a. Patch antenna is designed on the Rogers RT/duroid 5880 substrate with dielectric constant of 2.2 and thickness of 110 micrometers. The overall rod is surrounded by a cylindrical waveguide of height 3.5 mm as shown in Fig. 2b. This waveguide conducts the electromagnetic energy between the patch antenna and the dielectric rod antenna.

## C. Four-Elements Array:

For increasing gain, 4-elements array is proposed. The space between array elements is optimized to be 7 mm as shown in Fig. 3a. This maximizes the gain to 21 dBi.

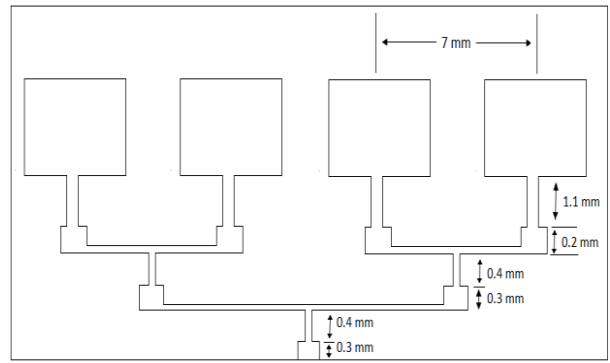

(a)

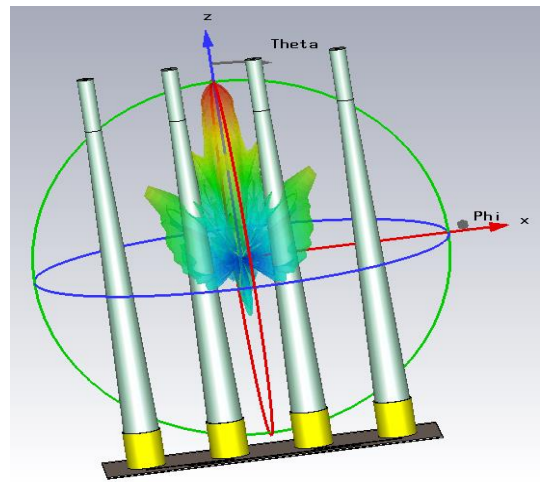

(b)

Fig. 3. (a) 4-patchs array, (b) 4-rods array.

## IV. SIMULATION RESULTS

### A. Single Dielectric Rod:

The structure was simulated using CST software. The optimized values of the dimensions are shown in Fig. 1 a, b and Fig. 2 a. The patch antenna itself has low gain and narrow bandwidth. By adding the solid dielectric rod, the resulting gain is increased from 7.86 to 16.8 dBi and bandwidth is increased from 1.019 to 2.35 GHz. By replacing the uniform solid sections with hollow uniform parts, the impedance bandwidth is increased up to be 4.4 GHz approximately. Fig. 4a shows the resulting $S_{11}$-Parameter for single patch, solid rod and hollow rod. Simulated Radiation pattern for single hollow rod is shown Fig. 4b.

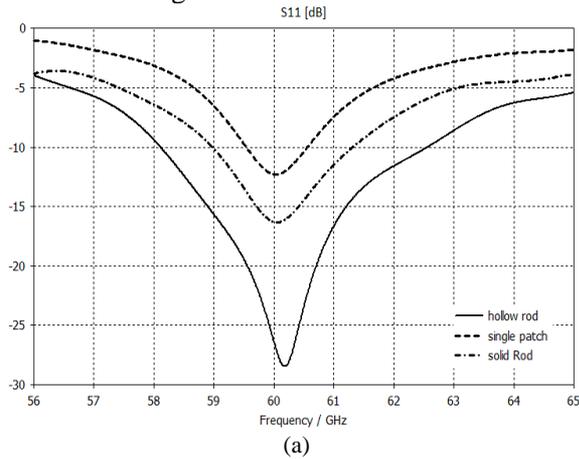

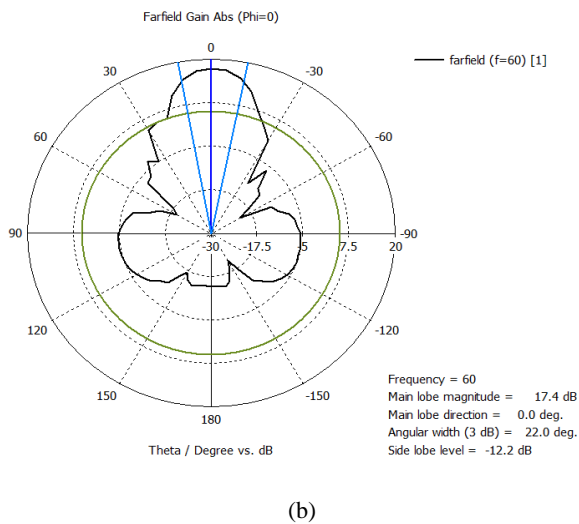

Fig. 4. (a) $S_{11}$ comparison for single patch, solid dielectric rod and hollow dielectric rod, (b) radiation pattern for hollow rod.

Fig. 5a compares the gain pattern of the patch antenna with a rod and without a rod. It is shown that, by adding a rod on the patch, the half power beamwidth (HPBW) is reduced from 76.4° to 22° and the gain increases up to 17.5 dB in a frequency bandwidth of 4.4 GHZ from 58 to 62.4 GHz.

The compromise between gain and HPBW can be obtained by the proper choice of the design parameters, namely the rod height, the waveguide height and the rod position with respect to the patch. Fig. 5b shows the radiation pattern at different height of the tapered section for the dielectric rod antenna.

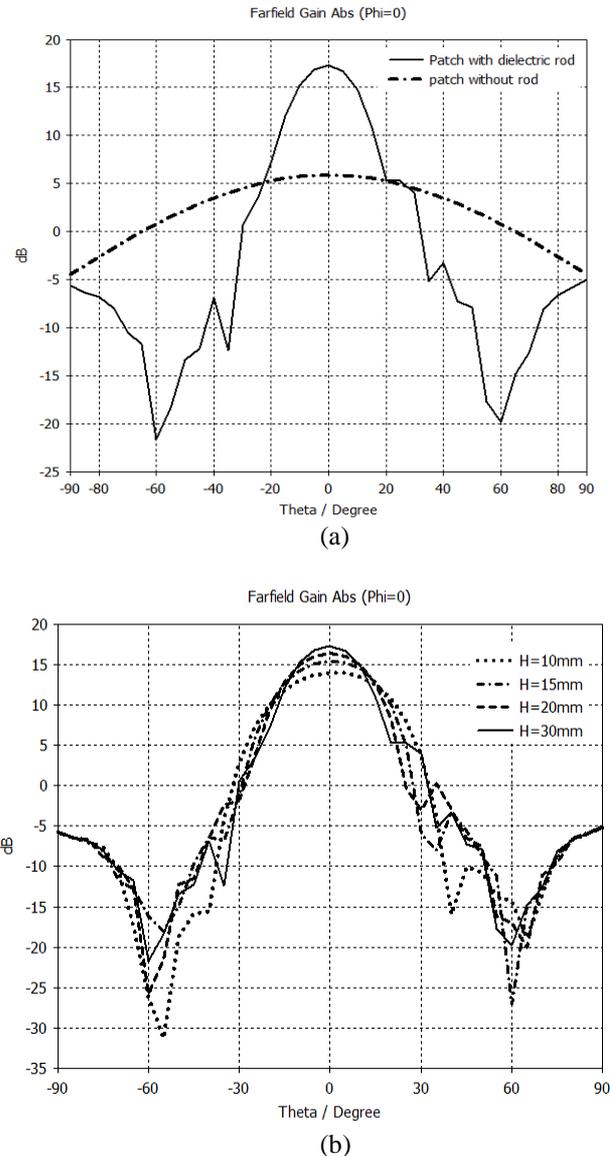

Fig. 5. Radiation pattern (a) for patch without/with dielectric rod, (b) at different height (H) of tapered section of Dielectric rod.

It is shown that the radiation pattern depends on the rod height. As the height increases, the gain increases and the HPBW reduces. But as the height increases above 30 mm, the antenna gain is saturated at 17.5 dBi.

When fixing the cylindrical rod height at 3 mm, Table I shows that higher gain can be achieved by increasing the height of the tapered rod.

Table I. Measured gain/HPBW at different height of tapered section of dielectric rod antenna.

| Height of $L_2$ | $2\lambda_0$ | $3\lambda_0$ | $4\lambda_0$ | $5\lambda_0$ | $6\lambda_0$ | $7\lambda_0$ | $8\lambda_0$ |
|---|---|---|---|---|---|---|---|
| Gain [dBi] | 14 | 15.4 | 16.4 | 16.9 | 17.5 | 17.5 | 17.5 |
| HPBW [deg] | 40.6 | 32.8 | 28.2 | 24.1 | 22.1 | 20.8 | 19.7 |

### *B. Four-Rods Array:*

Fig. 6 shows the resulting $S_{11}$-parameters for 4-patchs array, 4-solid rods and 4-hollow rods array. To improve the side lobe level, the inner diameter of the lower hollow rod is modified to $D_4=0.12\lambda_0$. This modification changes the side lobe level from -7 dB to -10.3 dB, but on the other hand, it decreases the bandwidth from 4 GHz to 3.82 GHz. It is clear that this proposed antenna resonates at two frequencies 58.89 GHz and 61.34 GHz. At frequency 58.89 GHz the gain of the antenna is 21.1 dB and at frequency 61.34 GHz the gain of the antenna is 21.4 dB.

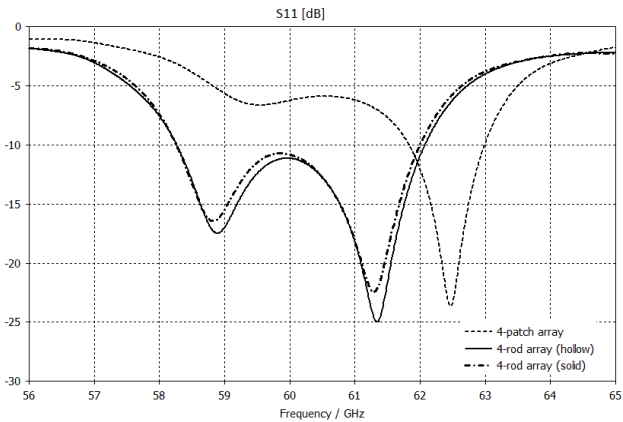

Fig. 6. $S_{11}$ comparison for 4-patches array, 4-solid rods and 4-hollow rods.

Fig. 7 (a and b) shows radiation pattern for 4-hollow rods array at 58.89 GHz and 61.34 GHz respectively.

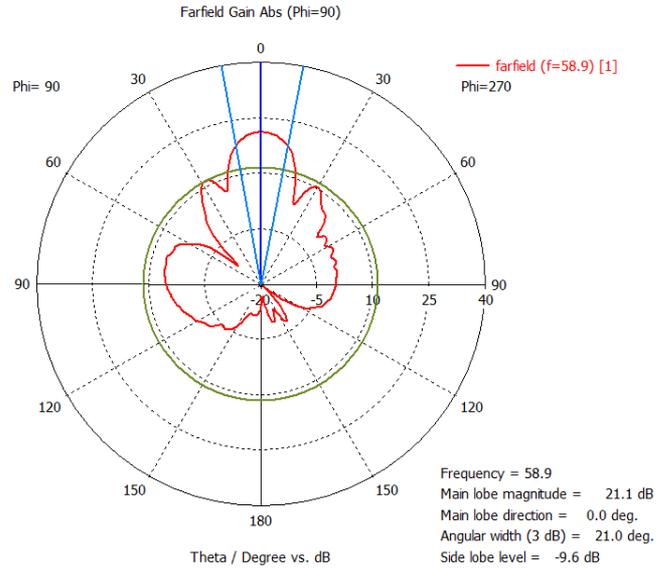

(a)

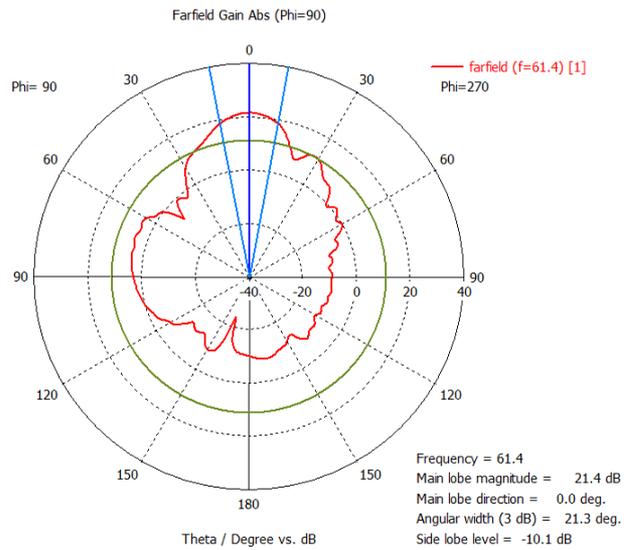

(b)

Fig. 7. Radiation pattern for 4-hollow rods at (a) 58.9 GHz (b) 61.4 GHz.

Table II presents simulated parameters comparison between single rectangular patch, solid rod, hollow rod and 4-hollow rods array. Results show that the increase in gain is on the price of increasing the side lobe level.

Table II. Comparison between single patch, dielectric rod and hollow dielectric rod.

|  | Single Patch | Solid Single rod | Hollow single rod | 4- hollow rod array |
|---|---|---|---|---|
| Gain [dBi] | 7.56 | 16.8 | 17.1 | 21.4 |
| Directivity [dBi] | 8.09 | 17.1 | 17.3 | 21.8 |
| SLL[dB] | -22.3 | -12.9 | -10.8 | -10.3 |
| BW[GHz] | 1.019 | 2.35 | 4.462 | 3.82 |

## V. CONCLUSION

In this paper, a new configuration of a tapered dielectric rod antenna fed by a patch is designed and discussed. This particular configuration is capable of producing higher gain for a given antenna length compared to the conventionally used tapered dielectric antennas. The hollow cylindrical section increases the bandwidth. Results show that we can obtain gain up to 17.5 dB and bandwidth of 4.4 GHz for single rod. For the 4-elements array, the resulting gain is around 21 dB and bandwidth is 3.82 GHz by optimizing the space between array elements and the hollow part diameter. The antenna can be used for high gain and high data rate applications especially dual-band applications. The proposed structure is very simple and suitable for consumer devices.

## REFERENCES

[1] M. Al Henawy and M. Schneider, "Planar antenna arrays at 60 GHz realized on a new thermoplastic polymer substrate," in 2010 Proceedings of the Fourth European Conference on Antennas and Propagation (EuCAP), April 2010, pp. 1 –5.

[2] N. Caillet, C. Person, C. Quendo, J. Favennec, S. Pinel, E. Rius, and J. Laskar, "High gain conical horn antenna integrated to a planar substrate for 60 GHz WPAN applications," in 2010 Proceedings of the Fourth European Conference on Antennas and Propagation (EuCAP), April 2010, pp. 1 –5.

[3] T. Ando, J. Yamauchi, and H. Nakano, "Numerical analysis of a dielectric rod antenna—Demonstration of the discontinuity-radiation concept," IEEE Trans. Antennas Propag., vol. 51, no. 8, pp. 2007–2013, Aug. 2003.

[4] R. E. Collin, F. J. Zucker, "Antenna Theory," McGraw Hill, Part 2,1969.

[5] R. Ala, R. Sadeghzadeh, and R. Kazemi, "Two-layer dielectric rod antenna for far distance," Antennas and Propagation Conference (LAPC), Loughborough, 2010.

[6] S. M. Hanham, T. S. Bird, A. D. Hellicar, and R. A. Minasian, "Optimized dielectric rod antennas for terahertz applications," 34th International Conference on Infrared, Millimeter, and Terahertz Waves (IRMMW-THz), 2009.

[7] Kumar, V. V. Srinivasan, V. K. Lakshmeesha, and S. Pal, "Design of short axial length high gain dielectric rod antenna," IEEE Transaction on Antennas and Propagation, Vol. 58, No. 12, Dec. 2010.

[8] K. C. Huang and D. J. Edwards, Millimetre wave antennas for gigabit wireless communications: A practical guide to design and analysis in a system context, 1st edition, Wiley, 2008.